\documentclass[epj]{webofc}
\usepackage[utf8]{inputenc}
\usepackage[varg]{txfonts}   
\usepackage{booktabs}
\usepackage{xcolor}
\definecolor{darkred}{rgb}{0.4,0.0,0.0}
\definecolor{darkgreen}{rgb}{0.0,0.4,0.0}
\definecolor{darkblue}{rgb}{0.0,0.0,0.4}
\usepackage[bookmarks,linktocpage,colorlinks,
    linkcolor = darkred,
    urlcolor  = darkblue,
    citecolor = darkgreen]{hyperref}
\usepackage{textcomp}
\usepackage{gensymb}
\usepackage{dsfont}
\wocname{EPJ Web of Conferences}
\woctitle{Lattice2017}
\newcommand{\tr}{\operatorname{tr}}

\newcommand{\corr}{\mathcal{C}}
\newcommand{\ssb}{s\bar{s}}
\newcommand{\lsb}{\ell\bar{s}}
\newcommand{\slb}{s\bar{\ell}}
\newcommand{\llb}{\ell\bar{\ell}}
\begin{document}
%
\selectlanguage{english}
\title{%
$\eta$ and $\eta^\prime$ masses and decay constants
}
\author{%
\firstname{Gunnar} \lastname{Bali}\inst{1,2} \and
\firstname{Sara}  \lastname{Collins}\inst{1} \and
\firstname{Jakob} \lastname{Simeth}\inst{1}\fnsep\thanks{Speaker, \email{jakob.simeth@ur.de}}
}
\institute{%
  Institute for Theoretical Physics, Universität Regensburg, D-93040 Regensburg, Germany
\and
  Department of Theoretical Physics, Tata Institute of Fundamental Research,
  Homi Bhabha Road, Mumbai 400005, India
}
\abstract{%
  We present preliminary results for the masses and decay constants of the
  $\eta$ and $\eta^\prime$ mesons using CLS $N_f = 2+1$ ensembles.

  One of the major challenges in these calculations are the large statistical
  fluctuations due to disconnected quark loops. We tackle these by employing
  a combination of noise reduction techniques which are tuned to minimize the
  statistical error at a fixed cost.

  On the analysis side we carefully assess excited states contributions by
  using a direct fit approach.
}
\maketitle
\section{Introduction}
\label{sec:org12c2077}
The \(\eta\) and \(\eta^\prime\) play an important role in 
many QCD processes and are directly connected to the chiral anomaly of QCD.
Their electromagnetic transition form factors are of great phenomenological interest 
and have been studied theoretically, e.g., using light cone sum rules \cite{Agaev:2014wna}. 
However, little is known about their wavefunctions and the quark mass dependence
of singlet-octet mixing and first-principles determinations are needed.
On the lattice, the \(\eta\) and \(\eta^\prime\) mesons are difficult to study.
Nonetheless, quite some progress has been made, e.g., in~\cite{Kuramashi:1994aj,Struckmann:2000bt,McNeile:2000hf,Lesk:2002gd,Christ:2010dd,Dudek:2011tt,Gregory:2011sg,Michael:2013vba,Bali:2014pva}.

The fact that the \(\eta\) and \(\eta^\prime\)
states are not flavour eigenstates presents a challenge. The mass
eigenstates are mixtures of flavour singlet and octet states, i.e.,
\begin{equation}
\mid \eta \rangle = \cos\theta \mid \eta_8 \rangle - \sin\theta \mid \eta_1\rangle\text{,}\qquad
\mid \eta^\prime \rangle = \sin\theta^\prime \mid \eta_8 \rangle + \cos\theta^\prime \mid \eta_1\rangle\text{,}
\end{equation}
where the angles \(\theta\) and \(\theta^\prime\) will in general depend on the
choice of \(\mid \eta_1\rangle\) and \(\mid \eta_8\rangle\), e.g., on the scale
and the smearing applied.

This can also be expressed in matrix form,
\begin{equation} \label{eq:gevp} 
\begin{pmatrix} \corr_\eta & 0 \\ 0 & \corr_{\eta^\prime} \end{pmatrix} = 
U \begin{pmatrix} \corr_{88} & \corr_{81} \\ \corr_{18} & \corr_{11} \end{pmatrix} U^T\text{, where}\quad
\corr_{ij} = \langle P_i(x) \mid P_j(y) \rangle\text{,}\quad\text{and}\quad
U(\theta, \theta^\prime) = \begin{pmatrix} \cos\theta & -\sin\theta \\ \sin\theta^\prime & \phantom{-}\cos\theta^\prime\end{pmatrix}
\end{equation}
is a non-unitary rotation matrix to allow for possible gluonic and other excited
states contributions.
$\corr_{ij}$ with \(i,j \in \{8,1\}\) are correlators connecting pseudoscalar octet and 
singlet operators, respectively:
\begin{equation}\label{eq:interpolators}
P_8 = \frac{1}{\sqrt{6}} \left(\bar{u}\gamma_5u + \bar{d}\gamma_5 d - 2 \bar{s}\gamma_5s\right) \quad \text{and}\quad
P_1 = \frac{1}{\sqrt{3}} \left(\bar{u}\gamma_5u + \bar{d}\gamma_5 d +   \bar{s}\gamma_5s\right)\text{.}
\end{equation}
When performing the Wick contractions for the matrix elements \(\corr_{ij}\), another
difficulty emerges: disconnected loops appear which are inherently noisy and
expensive to compute on the lattice. Efficient solvers for the Dirac equation and 
noise reduction techniques are mandatory to obtain decent signals.

In these proceedings we attempt a determination of the masses and leading distribution amplitudes
of the \(\eta\) and \(\eta^\prime\) mesons, refining our techniques on two of the many existing
\(N_f = 2+1\) CLS ensembles~\cite{Bruno:2014jqa} with non-perturbatively
improved Wilson fermions at $\beta=3.4$, 
corresponding to a lattice spacing \(a\approx 0.0864\,\mathrm{fm}\)
(determined using \(\sqrt{8\,t_0}=0.415\,\mathrm{fm}\) \cite{Bruno:2016plf}).
The first of the two, ensemble U103, is at the \(N_f=3\) symmetric point with
\(m_{\pi} = m_K \approx 415\,\textrm{MeV}\). The other, H105, is along the same line
of constant average quark mass, with \(m_\pi \approx 287\,\textrm{MeV}\)
and \(m_K \approx 487\,\textrm{MeV}\).

\section{Computation of disconnected loops}
\label{sec:org892e6b4}
The entries \(\corr_{ij}\) of the matrix in Eq. \eqref{eq:gevp} are given by
\begin{align}
\corr_{88} =& \frac{1}{3}\left(C_{\llb} + 2 C_{\ssb} - 2 D_{\llb} + 2 D_{\lsb} + 2 D_{\slb} - 2 D_{\ssb}\right)\label{eq:c88}\\
\corr_{18} = \corr_{81} =& \frac{\sqrt{2}}{3}\left(C_{\llb} - C_{\ssb} - 2 D_{\llb} + 2 D_{\lsb} - D_{\slb} + D_{\ssb}\right)\label{eq:c18}\\
\corr_{11} =& \frac{1}{3}\left(2 C_{\llb} + C_{\ssb} - 4 D_{\llb} - 2 D_{\lsb} - 2 D_{\slb} - D_{\ssb}\right)\label{eq:c11}\text{,}
\end{align}
and are linear combinations of connected and disconnected pseudoscalar correlators,
\begin{align}
C^{P}_{f\bar{f}}(\delta t) = \tr\left(\gamma_5 S^{-1}_{f}(\delta t) \gamma_5 S^{-1}_{f}(0)\right)\text{,}\\
D^{P}_{f\bar{g}}(\delta t) = \tr\left(\gamma_5 S^{-1}_{f}(\delta t)\right) \tr \left(\gamma_5 S^{-1}_{g}(0)\right)\text{,}
\end{align}
with light and strange quark flavours \(f,g \in \{\ell, s\}\) and \(S^{-1}_{f}\) being the quark propagator
of flavour \(f\). While the connected correlators can be cheaply computed by exploiting the \(\gamma_5\text{-hermiticity}\)
of the Dirac operator (we use one point source per configuration), the disconnected loops \(L_f(t) = \tr \gamma_5 S^{-1}_{f}(t)\) must be
estimated stochastically. We do so by inverting the Dirac equation on
time-diluted \(\mathbb{Z}_2 \otimes i \mathbb{Z}_2\) sources,
where a distance of \(4a\) between non-zero time slices was found to be optimal. 
Additionally, we employ the hopping parameter expansion to improve the signal-to-noise ratio of the loops
by exploiting \(L_f(t) = \tr \Gamma S^{-1}_{f}(t) = \kappa_f^k \tr \Gamma D^k
S^{-1}_{f}(t)\), where \(D\) here is the hopping part in the Dirac operator \(S_f = \mathds{1} + \kappa_f D\) and
\(k=2\) for \(\Gamma=\gamma_5\) and \(k=4\) for \(\Gamma=\gamma_\mu \gamma_5\)
are the maximum number of applications that can be used for our action.
See \cite{Bali:2009hu} and references therein for more details on these noise reduction techniques.

It turned out that 96 stochastic estimates for each quark flavour and time
dilution are also sufficient to 
obtain reasonable signals for the axialvector loop, that is needed for the extraction of the decay constants.
When performing this many solves per configuration, it is crucial to employ a
modern solver.
We use a multigrid solver that has been optimized specifically for the KNL architecture~\cite{qp3}.

\section{Extraction of physical states}
\label{sec:org771f516}
At the \(N_f = 3\) flavour symmetric point where \(S^{-1}_\ell = S^{-1}_s\) the elements of the correlator matrix,
Eqs.~\eqref{eq:c88} to \eqref{eq:c11}, simplify leading to a diagonal matrix: In this limit, the
\(\eta\) is a pure octet state and equal in mass to the pion, whereas the \(\eta^\prime\) is entirely
a singlet:
\begin{align}
  \corr_{88}(\delta t) = & \corr_{\eta}(\delta t) = \corr_\pi(\delta t),\\
  \corr_{11}(\delta t) = & \corr_{\eta^\prime}(\delta t) = \corr_\pi(\delta t) - 3 D(\delta t)\text{.}
\end{align}
For sufficiently large times, where excited states can be neglected, 
both the pion and the \(\eta^\prime\) correlators are single-exponentials.
Consequently, the disconnected correlator is a double-exponential
\begin{equation}\label{eq:disconfitformsympt}
D(\delta t) = \frac{1}{3} \left(\corr_\pi(\delta t) - \corr_{\eta^\prime}(\delta t)\right) 
\to \frac{1}{3}\left(A_\pi\exp(-m_\pi\delta t) - A_{\eta^\prime} \exp(-m_{\eta^\prime} \delta t)\right) \text{,}
\end{equation}
where \(A_\pi, A_{\eta^\prime} > 0\).
The (lighter) pion mass dominates at large times over the heavy
$\eta^\prime$,
which leaves only a small window for the extraction of the \(\eta^\prime\) state:
at very small times excited states impede the extraction of the ground state
while at large 
time separations the relative error arising from the disconnected diagrams grows rapidly and makes
resolving the small difference to the pion correlator impossible.
\begin{figure}[thbp]
\centering
  \includegraphics[width=.49\linewidth]{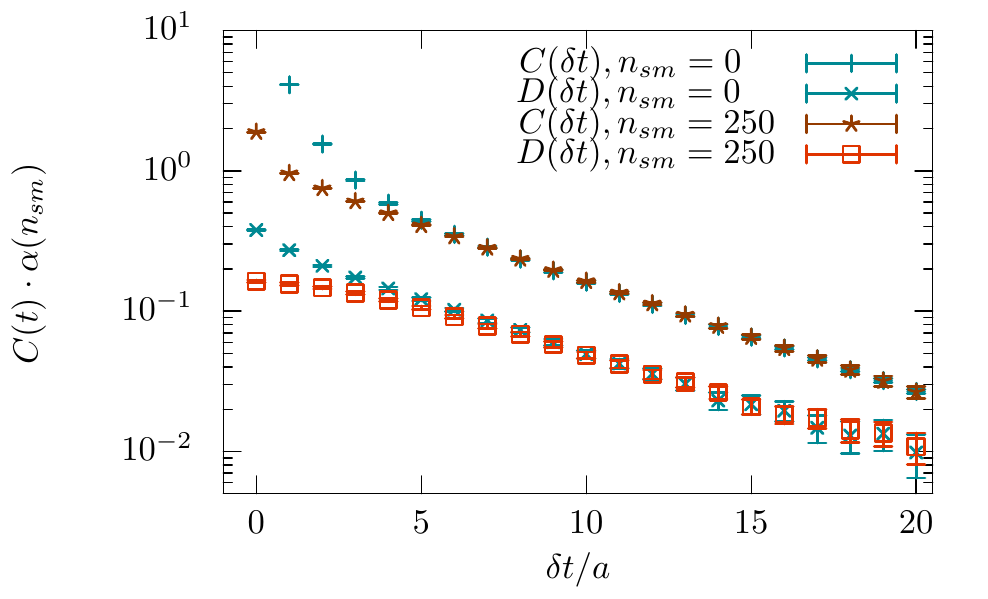}
  \hfill
  \includegraphics[width=.49\linewidth]{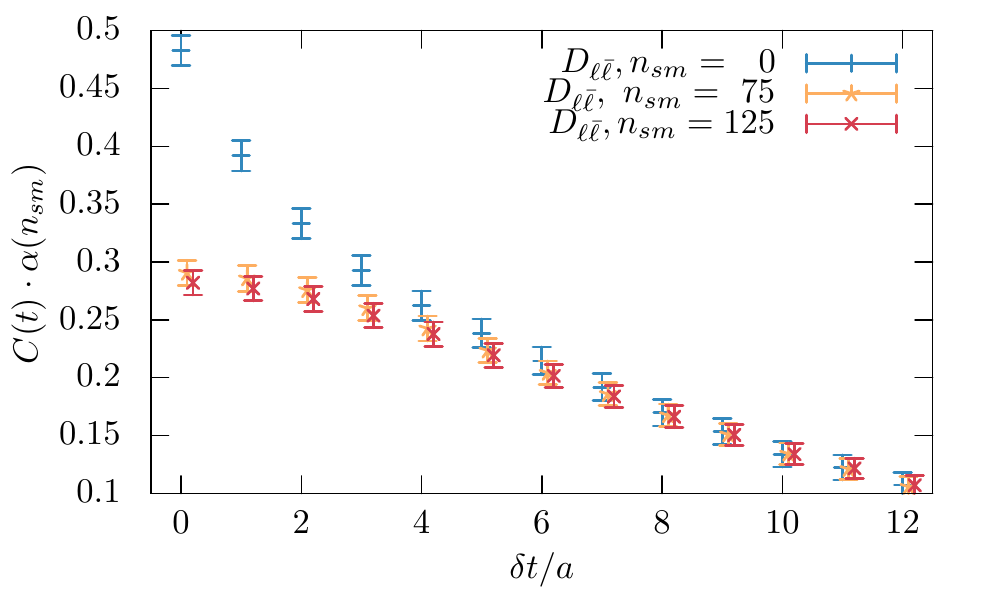}
  \caption{\label{fig:excStates}
    \textbf{left:} Pseudoscalar connected and disconnected correlators are
    shown, both for local sources and sinks ($n_{sm} = 0$) and after $n_{sm} =
    125$ smearing iterations for U103.
    The correlators are scaled to match such that they agree at large times.
    With no smearing excited states are clearly visible both in the connected
    and the disconnected correlators.
    \textbf{right:} The same for the light-light disconnected correlator of H105. The smeared
    correlators do not exhibit significant excited states after $n_{sm} = 75$ iterations.}
\end{figure}

We employ Wuppertal quark smearing \cite{Gusken:1989ad} both to the source and the sink loops to enhance the
ground over excited states. A small number of smearing iterations turned out to be sufficient
to suppress excited states and achieve moderate errors in the final results. In
particular for the disconnected correlators comparably few iterations seem to be
sufficient, see Fig.~\ref{fig:excStates}.

The observations above are also valid away from the symmetric point. There, the correlator matrix,
Eq. \eqref{eq:gevp}, must first be diagonalized to extract the physical states. This is usually done
by solving the generalized eigenvalue problem, yielding the physical correlators as eigenvalues.
However, we observed comparably large errors and unstable results when changing 
the parameters of the method. Instead, we insert Eqs.~\eqref{eq:c88}~--~\eqref{eq:c11}
into Eq.~\eqref{eq:gevp} and obtain a system of three linear
equations that we solve for the three independent disconnected correlators, \(D_{\llb},
D_{\lsb}, \text{and} D_{\ssb}\), yielding similar expressions as at the symmetric point, Eq.~\eqref{eq:disconfitformsympt}. 
In the \(N_f=2+1\) case, the resulting formulae are more complex
and depend on two angles \(\theta\) and \(\theta^\prime\), two connected correlators
for the pion and the \(s\bar{s}\), and the wanted \(\eta\) and \(\eta^\prime\) correlators. Plugging
single-exponentials into the correlators, we end up with a total of ten independent parameters.

\begin{figure}[thbp]
  \centering
  \includegraphics[width=.49\linewidth]{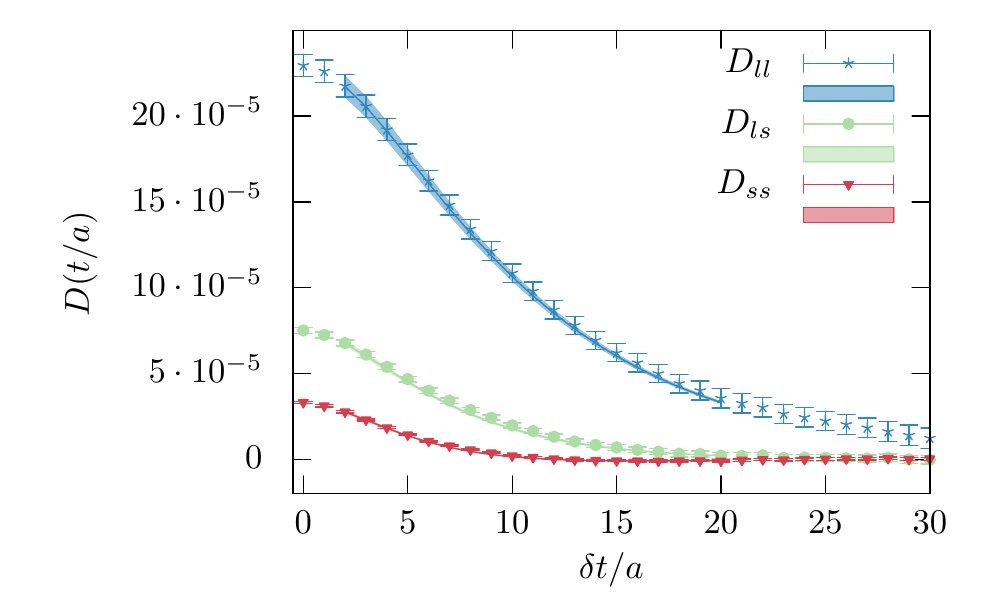}
  \hfill
  \includegraphics[width=.49\linewidth]{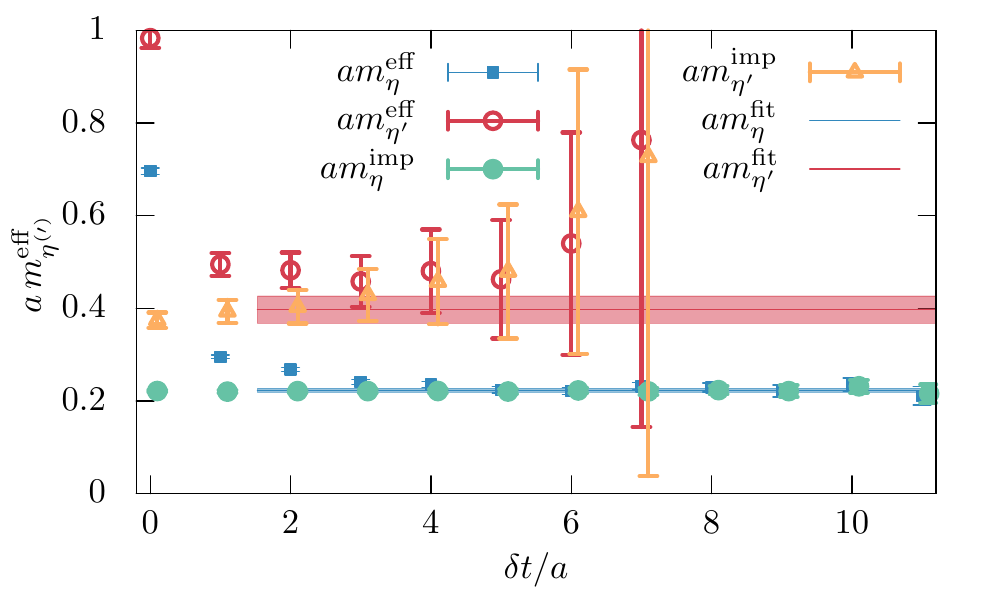}
  \caption{\label{fig:fits}
    \textbf{left:}~Combined fit to the disconnected correlators of H105, using information from
    a previous fit to the connected correlators. We obtain typical values of
    \(\chi^2/\mathrm{d.o.f.} \approx 1\) in correlated fits.
    \textbf{right:}~Effective masses for the \(\eta\) and \(\eta^\prime\),
    computed from their reconstructed correlators, denoted
    \(a\,m^{\mathrm{eff}}_{\eta^{(\prime)}}\), and with the connected
    correlators replaced by their fitted curves (\(a\,m^{\mathrm{imp}}_{\eta^{(\prime)}}\)), along with the fitted mass
    values from the combined fit. Shaded regions represent the errors of the
    fitted masses and the fit range of the disconnected correlators. Connected
    correlators are fitted in the range \(\delta t/a \in [15,40]\). Excited
    states arising from the connected correlators at earlier times are
    clearly visible.
  }
\end{figure}
We perform a combined fit to determine these parameters. The connected
correlators naturally do not depend on the disconnected ones and can be fitted first, reducing the number 
of parameters to six. Being able to tune the fit ranges independently we can fit the connected
correlators in a region where we are sure to be free of excited states contaminations whereas
the disconnected fits can start at smaller times to capture the heavy \(\eta^\prime\) state sufficiently
well. At later times, the well determined connected correlators constrain the fit.
This approach is similar in spirit to the subtraction method of~\cite{Neff:2001zr,Jansen:2008wv,Michael:2013vba}.
Fig.~\ref{fig:fits} demonstrates that the fit describes
the data points and the fit errors actually decrease at large times, confirming
our expectations.

Note that in Eq. \eqref{eq:gevp} we chose the octet-singlet basis, which is just one particular choice
and we can as well work in any other basis, e.g., in the flavour basis, where
\begin{align}
\mid \eta \rangle        = \frac{1}{\sqrt{2}} \cos\vartheta        \mid \bar{u}\gamma_5 u + \bar{d}\gamma_5d\rangle - \sin\vartheta        \mid \bar{s}\gamma_5s\rangle\\
\mid \eta^\prime \rangle = \frac{1}{\sqrt{2}} \sin\vartheta^\prime \mid \bar{u}\gamma_5 u + \bar{d}\gamma_5d\rangle + \cos\vartheta^\prime \mid \bar{s}\gamma_5s\rangle\text{,}
\end{align}
with different angles \(\vartheta^{(\prime)}\). In this case, the fit forms look a bit simpler but
the resulting correlators and masses are (numerically) equal.

To remove any potential constant shifts from our (disconnected) correlators stemming from an 
incomplete sampling of the topological sectors, we also fit to
\begin{equation}
 - \hat{\partial}_4 \corr(\delta t) \approx \frac{\corr(\delta t - a) - \corr(\delta t + a)}{2a}
\end{equation}
as suggested in \cite{Feng:2009ij,Umeda:2007hy}. We found that this reduces the error and also
stabilizes the result with respect to different fit ranges.
%

In order to further improve statistics in the mass estimates, and as a cross check on systematic
errors we also combine data at finite momentum and fit to the dispersion relation
\begin{equation}\label{eq:disprel}
a E(m, p) = \sqrt{a^2 m^2 + a^2 p^2}\text{,}
\end{equation}
where \(a^2p^2 = \sum_\mu k^2_\mu \frac{4 \pi^2}{L_\mu^2}\) is the square of the lattice momentum,
\(L_\mu\) is the number of lattice points in direction \(\hat{\mu}\) and \(-L_\mu/2 \leq k_\mu < L_\mu/2\)
are the integer momentum components.
\begin{figure}[thbp]
\centering
  \includegraphics[width=.49\linewidth]{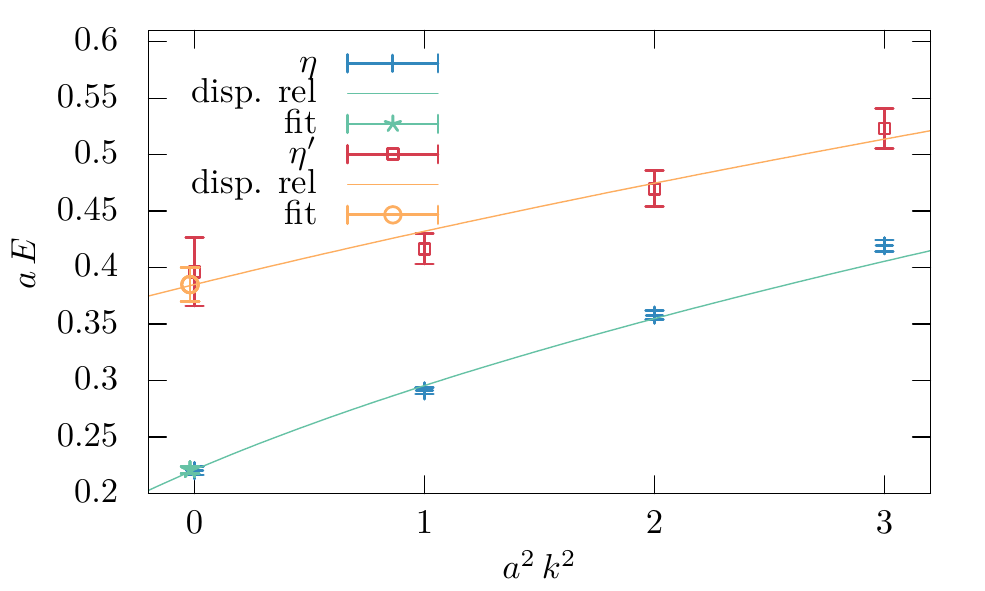}
\hfill
  \includegraphics[width=.49\linewidth]{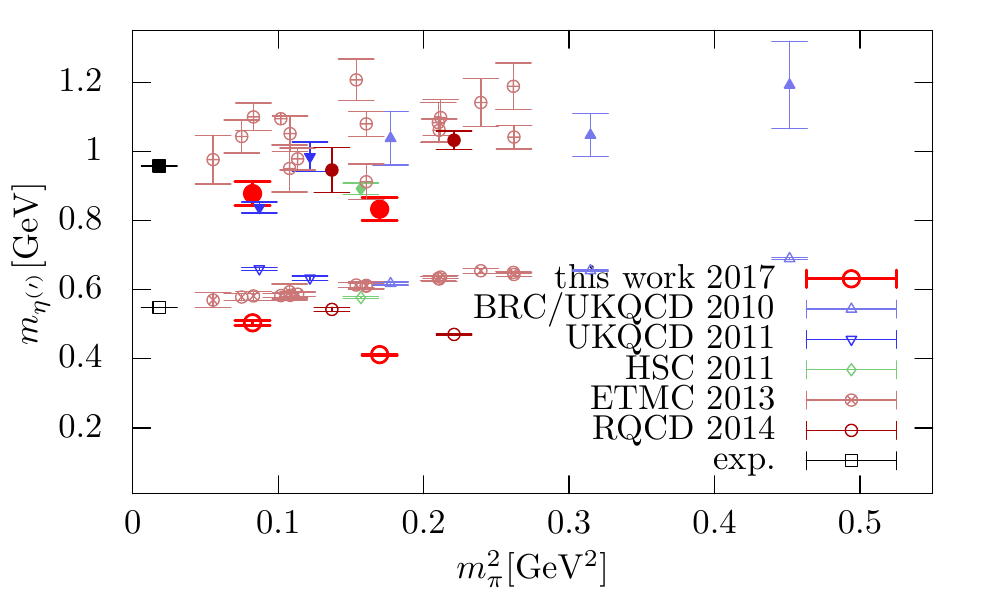}
  \caption{\label{fig:masssummary}
    \textbf{left:} Fits to the dispersion relation, Eq.~\eqref{eq:disprel}, combining data at several
    momenta.
    \textbf{right:} Summary plot of recent lattice determinations of the
    \(\eta\) and \(\eta^\prime\) masses,
    cf.~\cite{Christ:2010dd,Gregory:2011sg,Dudek:2011tt,Michael:2013vba}.}
\end{figure}
From this fitting approach, we obtain for the U103 ensemble at the flavour symmetric point,
\begin{equation}
m_\eta          = m_\pi = 412(2)\,\mathrm{MeV},\quad m_{\eta^\prime} = 833(33)\,\mathrm{MeV}\text{,}
\end{equation}
and away from that point, for H105, at a pion mass of \(282\,\mathrm{MeV}\),
\begin{equation}
m_\eta          = 504(7)\,\mathrm{MeV},\quad m_{\eta^\prime} = 878(35)\,\mathrm{MeV}\text{.}
\end{equation}
Fig. \ref{fig:masssummary} compares these results with other lattice determinations and the physical point.
It should be stressed that, like in the RQCD 2014 study~\cite{Bali:2014pva}, the
two considered ensembles are on a line of constant average quark mass and
approach the physical point on a different trajectory than other (\(m_s
\approx \mathrm{const.}\)) studies.

\section{Decay constants}
\label{sec:org08edb4a}
Having extracted masses and the angles, we can construct the correlators
\begin{equation}
\corr_{A^i\eta^{(\prime)}}(\delta t) = \langle 0 \mid A^i_\mu(\delta t) \mid \eta^{(\prime)}\rangle\text{,}\quad
\corr_{P^i\eta^{(\prime)}}(\delta t) = \langle 0 \mid P^i(\delta t) \mid \eta^{(\prime)}\rangle\rangle\text{,}\quad
\corr_{\eta^{(\prime)}\eta^{(\prime)}}(\delta t) = \langle \eta^{(\prime)}(\delta t) \mid \eta^{(\prime)}\rangle
\end{equation}
where \(P^i\) labels the octet (\(i=8\)) and singlet (\(i=1\)) pseudoscalar
local currents, defined in Eq.~\eqref{eq:interpolators}
and \(A^i_\mu\) similarly defines the axialvector currents.
Using these correlators, one can compute decay constants which read
\begin{equation}
\langle 0 \mid A_\mu^i \mid \eta^{(\prime)}(p) \rangle = i f_N^i p_\mu\text{.}
\end{equation}
This results in four different effective decay constants at momentum \(p=0\),
\begin{equation}\label{eq:effdecconst}
f^{i,\mathrm{eff}}_{\eta^{(\prime)}}(\delta t) = Z_A^i \frac{\sqrt{2}\corr_{A^i\eta^{(\prime)}}(\delta t)}{\sqrt{\corr_{\eta^{(\prime)}\eta^{(\prime)}}(\delta t) m_{\eta^{(\prime)}}}}\exp(m_{\eta^{(\prime)}} \delta t/2)\text{,}
\end{equation}
which are constant over some range in \(\delta t\) where excited states can be neglected and the correlators have not
yet vanished in the noise.

For (partial) \(\mathcal{O}(a)\) improvement, we replace
\begin{equation}
\corr_{A^i\eta^{(\prime)}} \to \tilde{\corr}_{A^i\eta^{(\prime)}} = (1+b_A a\,m_i) (\corr_{A^i\eta^{(\prime)}} + a c_A \partial_4 \corr_{P^i\eta^{(\prime)}})\text{,}
\end{equation}
where \(a\,m_8 = \frac{a}{3}(m_l + 2\,m_s)\) and \(a\,m_1 = \frac{a}{3}(2\,m_l + m_s)\).
We take the non-perturbatively determined improvement coefficients
\(b_A\) from~\cite{Korcyl:2016ugy}, \(c_A\) from~\cite{Bulava:2015bxa} and \(Z_A^8\) from~\cite{Bulava:2016ktf}. 
For the axial renormalization factor, the difference between \(Z_A^8\) and
\(Z_A^1\) is of \(\mathcal{O}(\alpha_s^2)\) and known perturbatively
\cite{Constantinou:2016ieh}.
For this lattice spacing and action it is at the level of 2 \%, which is in 
line with preliminary recent non-perturbative estimates \cite{Bali:2017jyw}.
From these numbers, we estimate the systematic uncertainty from renormalization
to be around 5 \%.
In this work, we take the non-perturbative octet renormalization factor together with the
perturbative difference at a scale of $2\,\mathrm{GeV}$.

\begin{figure}[thbp]
\centering
\includegraphics[width=.8\linewidth]{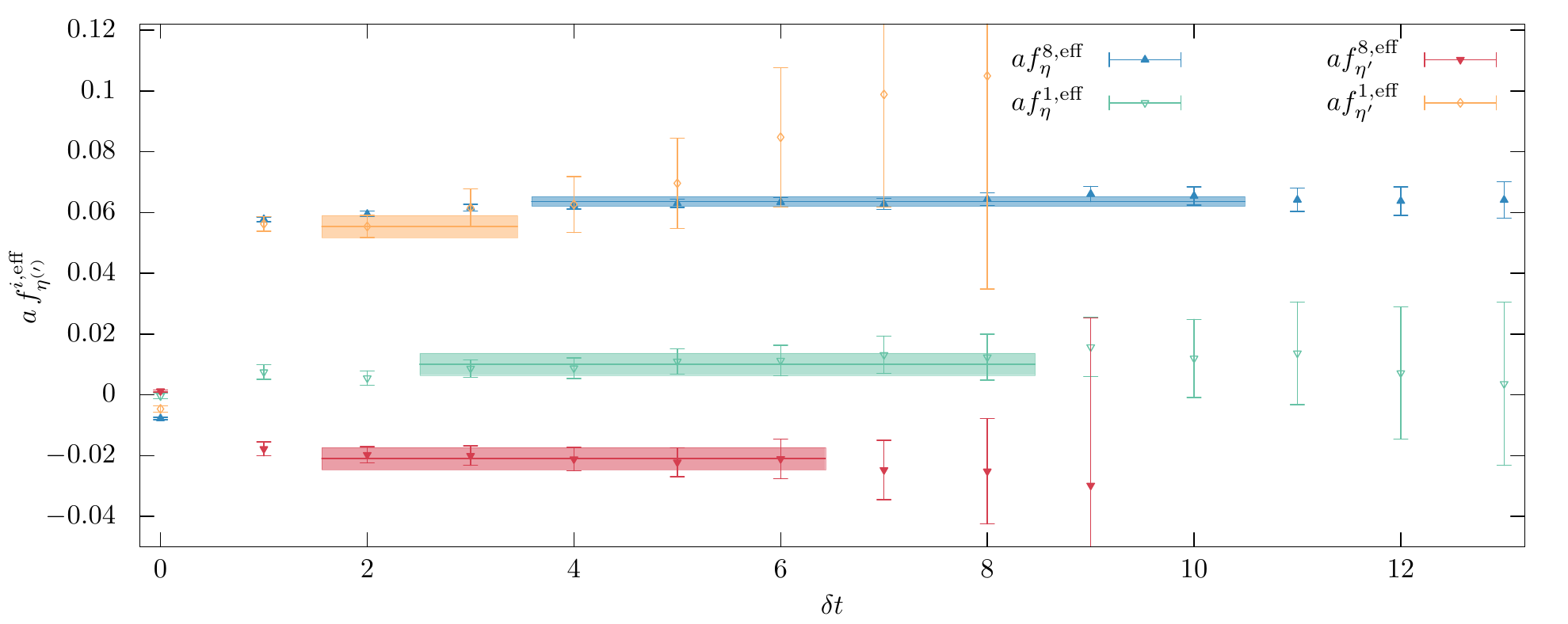}
\caption{\label{fig:effdecconst}
Effective decay constants as defined in Eq. \eqref{eq:effdecconst} for H105.
Shaded regions indicate fit ranges and errors.}
\end{figure}

As can be seen in Fig.~\ref{fig:effdecconst}, we face the "window problem" again: The axialvector loops are even noisier
than the pseudoscalar ones so that the signal is lost very early in Euclidean time.
Also, being local at the sink, we need to take special care of excited states. 
By fixing the masses to their previously determined values we can take weighted
averages ("one-parameter fits") to each of the effective decay constants, tuning
the ranges individually. The results are compatible with~\cite{Bali:2014pva},
however, \(f_{\eta^\prime}^1\) is quite a bit larger, which might be due to
insufficient treatment of excited states in that channel.

It is common to give the decay constants in the flavour basis, which we recover
from the fitted values by taking the linear combinations at the scale of
interest, \(\mu = 2\,\mathrm{GeV}\),
\begin{equation}
    f_{\eta^{(\prime)}}^l =   \sqrt{\frac{1}{3}} f_{\eta^{(\prime)}}^8 + \sqrt{\frac{2}{3}}f_{\eta^{(\prime)}}^1,\quad
    f_{\eta^{(\prime)}}^s = - \sqrt{\frac{2}{3}} f_{\eta^{(\prime)}}^8 + \sqrt{\frac{1}{3}}f_{\eta^{(\prime)}}^1\text{.}
\end{equation}
Usually, these are parameterized in terms of two constants and two angles,
\begin{equation}
\begin{pmatrix}f_\eta^l & f_\eta^s\\ f_{\eta^\prime}^l & f_{\eta^\prime}^s\end{pmatrix} = \begin{pmatrix}f_l\,\cos\theta_l & -f_s\,\sin\theta_s\\ f_l\,\sin\theta_l & f_s\,\cos\theta_s\end{pmatrix} = U^T(-\theta_l, -\theta_s) \begin{pmatrix}f^l & 0 \\ 0 & f^s\end{pmatrix}\text{,}
\end{equation}
which we extract simply via
\begin{align}
  \theta_l = &   \arctan\frac{f_{\eta^\prime}^l}{f_{\eta}^l       }\text{,}&\quad
  \theta_s = & - \arctan\frac{f_{\eta}^s       }{f_{\eta^\prime}^s}\text{,}&
   f_l = & \frac{f_{\eta}^l       }{2 \cos\theta_{l}} + \frac{f_{\eta^\prime}^l}{ 2 \sin\theta_{l}}\text{,}&\quad
   f_s = & \frac{f_{\eta^\prime}^s}{2 \cos\theta_{s}} - \frac{f_{\eta}^s       }{ 2 \sin\theta_{s}}\text{.}
\end{align}
Additionally, we monitor unitarity violations by considering the difference of
the angles \(\delta \theta = \theta_l - \theta_{s}\),
which has a non-vanishing value if there is a gluonic contribution to \(f_{\eta^{(\prime)}}^i\)
and is related to the low-energy constant \(\Lambda_1\) \cite{Feldmann:1999uf},
\begin{equation}
f_l f_s \sin(\theta_l - \theta_s) = \frac{\sqrt{2}}{3} f_\pi^2 \Lambda_1\text{.}
\end{equation}
For the two ensembles, we obtain for the decay constants
\begin{align}
\text{U103:}\quad f_l/f_\pi = & 1.20(7),&\quad  f_s/f_\pi = & 1.27(2)\text{,}\qquad
\text{H105:}\quad f_l/f_\pi = & 1.27(7),&\quad f_s/f_\pi = &  1.39(5)\text{,}
\end{align}
and for the two angles and \(\Lambda_1\)
\begin{align}
\text{U103:}\quad \theta_l = & 61.3(1.8)\degree,&\quad \theta_s = & 65.4(1.6)\degree,&\quad \delta \theta = & -4.1(3.4)\degree,&\quad \Lambda_1 = -0.23(17)\text{,}\\
\text{H105:}\quad \theta_l = & 59.7(1.6)\degree,&\quad \theta_s = & 66.9(1.3)\degree,&\quad \delta \theta = & -7.2(2.9)\degree,&\quad \Lambda_1 = -0.47(17)\text{,}
\end{align}
Within present statistics, we find \(\theta_l \approx \theta_s\), i.e., we do not see any strong OZI
violation, however, the difference increases at the H105 ensemble compared to the
symmetric point. It will be interesting to see if this becomes stronger at
smaller quark masses.

Compared to phenomenology, the light decay constants have rather large values:
\begin{align}
\text{\cite{Feldmann:1998vh,Feldmann:1998sh}}\quad f_l/f_\pi = & 1.07(2),&\quad  f_s/f_\pi = & 1.34(6)\text{,}&\quad\theta_l = \theta_s =39.3(1.0)\text{,}\\
\text{\cite{Escribano:2005qq}}\quad f_l/f_\pi = & 1.09(3),&\quad  f_s/f_\pi = & 1.66(6)\text{,}&\quad\theta_l = \theta_s=40.7(1.4)\text{.}
\end{align}
As already mentioned, this might be due to underestimated excited states, in particular in the
correlator~\(\corr_{A^i\eta^\prime}^1\). Being local at the sink, the chosen smearing
might be insufficient to completely remove them. It has also been shown
\cite{Bruno:2016plf} that
cutoff effects are large, e.g., for the combination \(\sqrt{8\,t_0}f_{\pi K}\) in particular at this
lattice spacing. It remains to be seen how the values change when
approaching the physical point, i.e., at finer lattices and smaller quark masses.

\section{Summary}
\label{sec:org8171312}
In these proceedings, we employed several noise reduction techniques to obtain
disconnected loops at a precision that allows us to study the \(\eta/\eta^\prime\)
system on two CLS ensembles.

To extract the physical states, we directly fit to the correlators.
This, in combination with quark
smearing enables us to capture the small-t behaviour, encoding the
$\eta^\prime$ physics and still have excited states under control.
In a second fit we incorporate even noisier data from the axialvector channel
to determine decay constants directly.
The results are encouraging, however, a detailed analysis of systematic errors
still remains to be done.

We plan to do so by analysing more CLS ensembles, following two distinct
quark mass trajectories. The combination of ensembles along the line of constant
average quark mass with ensembles where the strange quark mass is kept fixed, as
well as going towards lighter masses will allow for a controlled chiral
extrapolation. Also finite lattice spacing and volume effects will be investigated.
\section{Acknowledgments}
\label{sec:org358174c}
This work was supported by the DFG SFB/TRR 55.
We thank our colleagues in CLS [\url{http://wiki-zeuthen.desy.de/CLS/CLS}]
for the joint effort in the generation of the gauge field ensembles 
which form a basis for the here described computation. 
The authors gratefully acknowledge the Gauss Centre for Supercomputing (GCS) 
for providing computing time through the John von Neumann Institute for
Computing (NIC) on the GCS share of the super-computer JUQUEEN at Jülich 
Supercomputing Centre (JSC). GCS is the alliance of the three 
national supercomputing centers HLRS (Universität Stuttgart),
JSC (Forschungszentrum Jülich), and LRZ (Bayerische Akademie der Wissenschaften), 
funded by the German Federal Ministry of Education and Research (BMBF) and the 
German State Ministries for Research of Baden-Württemberg (MWK), Bayern (StMWFK)
and Nordrhein-Westfalen (MIWF).
\bibliography{lattice2017}

\end{document}